\documentstyle{article}

 \newtheorem{claim}{Claim}
 
 \newtheorem{theorem}{Theorem}

\newenvironment{proof}{\par \sf Proof.\rm}{\hspace*{\fill}$\Box$\vspace{1ex}}

\begin{document}


\title{A Lower Bound on the Average-Case Complexity of Shellsort}

\author{Tao Jiang
\thanks{
  Department of Computer Science, University of California,
  Riverside, CA 92521. Email: jiang@cs.ucr.edu. 
  Supported in part by NSERC and CITO grants, and a UCR startup grant.}
 \and 
Ming Li
\thanks{Dept of Computer Science, 
University of California, Santa Barbara, CA 93106, USA (on leave from
the University of Waterloo). 
Email: mli@cs.ucsb.edu. 
Supported in part by NSERC and CITO grants and Steacie Fellowship.}
 \and
Paul Vit\'{a}nyi
\thanks{CWI, Kruislaan 413, 1098 SJ Amsterdam,
The Netherlands.  Email: paulv@cwi.nl.
Supported in part by the EU via NeuroCOLT II Working Group and
the QAIP Project.}
}

\maketitle

\begin{abstract}
We demonstrate an  $\Omega (p n^{1+1/p})$
lower bound on the average-case running time (uniform distribution)
of $p$-pass Shellsort. This is the first nontrivial
general lower bound for average-case Shellsort.

Additional Key Words and Phrases: Sorting, Shellsort,
computational complexity, average-case complexity,  Kolmogorov complexity.
\end{abstract}

\section{Introduction}
The question of a nontrivial general lower bound 
(or upper bound) on the average complexity of Shellsort
(due to D.L. Shell \cite{Sh59})
has been open for about four decades \cite{Kn73,Se97}.
We present such a lower bound for $p$-pass Shellsort
for every $p$. 

Shellsort sorts a list of $n$ elements in
$p$ passes using a sequence of increments
$h_1 , \ldots , h_p$. In the $k$th pass the
main list is divided in $h_k$ separate sublists
of length $\lceil n/h_k \rceil$,
where the $i$th sublist consists of the elements at positions
$j$, where $j \bmod h_k = i-1$, of the main list ($i=1, \ldots , h_k$).
Every sublist is sorted using a straightforward insertion sort.
The efficiency of the method is governed
by the number of passes $p$ and
the selected increment sequence $h_1 , \ldots , h_p$
with $h_p =1$ to ensure sortedness of the final list.
The original $\log n$-pass 
\footnote{``$\log$'' denotes the binary logarithm and ``$\ln$''
denotes the natural logarithm.
}
increment sequence 
$\lfloor n/2 \rfloor , \lfloor n/4 \rfloor, \ldots , 1$ 
of Shell \cite{Sh59} uses worst case $\Theta (n^2)$ time,
but Papernov and Stasevitch \cite{PS65} showed that 
another related sequence uses $O(n^{3/2})$, and 
Pratt \cite{Pr72} extended this to a class of all nearly
geometric increment sequences and proved this bound 
was tight.
The currently best asymptotic method was found
by  Pratt \cite{Pr72}. It uses
all $\log^2 n$ increments of the form 
$2^i 3^j < \lfloor n/2 \rfloor$
to obtain time $O(n\log^2 n)$ in the worst
case. Moreover, since every pass
takes at least $n$ steps, the average complexity 
using Pratt's increment sequence is $\Theta (n \log^2 n)$.
Incerpi and Sedgewick \cite{IS85} constructed a family of
increment sequences for 
which Shellsort runs in $O(n^{1+ \epsilon / \sqrt{\log n}})$ time
using $(8/\epsilon^2) \log n$ passes,
for every $\epsilon > 0$.  B. Chazelle (attribution in  \cite{Se96})
obtained the same result by generalizing Pratt's method:
instead of using 2 and 3 to construct the increment sequence
use $a$ and $(a+1)$ for fixed $a$ which yields a 
worst-case running time of $n \log^2 n (a^2/\ln^2 a)$
which is $O(n^{1+ \epsilon / \sqrt{ \log n}})$ for $\ln^2 a = O(\log n)$.
Plaxton, Poonen and Suel \cite{PPS92} proved
an $\Omega (n^{1 + \epsilon / \sqrt{p}} )$
lower bound for $p$ passes of Shellsort using 
any increment sequence, for some $\epsilon > 0$; 
taking $p = \Omega (\log n)$ shows that the Incerpi-Sedgewick / Chazelle
bounds are optimal for small $p$ and taking $p$ slightly larger
shows a $\Theta ( n \log^2 n / (\log \log n)^2)$ lower bound
on the worst case complexity of Shellsort.
Since every pass takes at least $n$ steps this shows
an $\Omega (n \log^2 n/ (\log \log n)^2)$ lower bound
on the worst-case of every Shellsort increment sequence.
For the {\em average-case} running time Knuth \cite{Kn73} showed
$\Theta ( n^{5/3})$ for the best choice of increments in $p=2$ passes;
Yao \cite{Yao80} analyzed
the average case for $p=3$ but did  not obtain a simple analytic
form; Yao's analysis was improved by 
Janson and Knuth~\cite{JK96} who showed $O(n^{23/15})$ average-case 
running time for a particular choice of increments in $p=3$ passes. 
Apart from this no nontrivial results
\footnote{The trivial lower bound is $pn$ comparisons since every
element needs to be compared at least once in every pass.} 
are known for the average case; see \cite{Kn73,Se96,Se97}.

\bigskip
{\noindent \bf Results:} 
We show the result given in the abstract, more precisely,
Theorem~\ref{theo.main}. The main result
is Theorem~\ref{theo.shelllb} below.
This is the first
advance on the problem 
of determining general nontrivial bounds on the {\em average-case} running time
of Shellsort~\cite{Pr72,Kn73,Yao80,IS85,PPS92,Se96,Se97}. 
The proof was originally obtained using Kolmogorov complexity 
(for Kolmogorov complexity see \cite{LiVi93}). 
The idea is to consider
an ``individually random'' permutation of the input list (a permutation
incompressible in the sense of Kolmogorov complexity).
If one 
encodes every move of Shellsort cheaply,
and if the algorithm does not make a certain number of moves, 
then one obtains
a too short encoding of the random permutation---contradicting the
incompressibility of it. 
\footnote{Fix a Shellsort algorithm.
Code the lengths of the inversion insertion paths in appropriate fixed order.
Since the input permutation
can be reconstructed from the coding,  the overall length of
the code must exceed the length of the shortest description of the
input permutation. Since the latter is assumed to be incompressible this
gives a lower bound on the sum total of the lengths of insertion paths
and hence on the running time.}
Moreover, since the overwhelming 
majority of permutations is incompressible we obtain the bound
on the average. 
It turns out that the argument can
be translated to a counting argument. This we have done and present only
the more elementary and shorter counting argument here. 
The original proof using Kolmogorov complexity is given in the preliminary
version \cite{JLV99}.
It is instructive that thinking in terms of
code length and Kolmogorov complexity enabled advances in this
problem.

\section{The Lower Bound}
A Shellsort computation consists of a sequence of
comparison and inversion (swapping) operations. 
We count just
the total number of data movements (here inversions) executed.
The lower bound obtained below 
holds {\em a fortiori} for the number of comparisons.
The proof is based on the following intuitive idea:
There are $n!$ different permutations. Given the sorting process (the
insertion paths in the right order) one can recover the correct permutation
from the sorted list.
Hence one requires $n!$ pairwise different sorting processes. This gives 
a lower bound on the minimum of the maximal length of a process.

\begin{theorem}\label{theo.shelllb}
Let $0 < \epsilon < 1$ and $n,p$ satisfy $p \leq ( \epsilon \log n)/ \log e$.
For every $p$-pass Shellsort algorithm and every increment 
sequence, every subset of $n!/2^n$ 
input permutations of $n$ keys
contains an input permutation
that uses
$
\Omega \left( pn^{1+(1 - \epsilon)/p}
\right)$ 
inversions (and comparisons).
\end{theorem}

\begin{proof}
Let the list to be sorted consist of a permutation
$\pi$ of the elements $1, \ldots , n$.
Consider a $(h_1 , \ldots , h_p)$ Shellsort algorithm $A$ where
$h_k$ is the increment in the $k$th pass and $h_p=1$.
For any $1 \leq i \leq n$ and $1 \leq k \leq p$, let $m_{i,k}$ be 
the number of elements in the {\em $h_k$-chain} containing element $i$
that are to the left of $i$ at the beginning of pass $k$  
and are larger than $i$.
Observe that $\sum_{i=1}^{n} m_{i,k}$ is the number of inversions
in the initial permutation of pass $k$, and that the insertion sort
in pass $k$ requires precisely $\sum_{i=1}^{n} (m_{i,k} +1)$
comparisons.
Let $N$ denote the total number of inversions:
\begin{equation}\label{eq.M}
 N := \sum_{k=1}^p \sum_{i=1}^{n} m_{i,k}. 
\end{equation}

\begin{claim}\label{lem.descr}
Given all the  $m_{i,k}$'s in an appropriate fixed order,
we can reconstruct the original permutation $\pi$.
\end{claim}
\begin{proof}
The $m_{i,p}$'s trivially specify the initial permutation of pass $p$.
In general, given the $m_{i,k}$'s and the final permutation of pass $k$,
we can reconstruct the initial permutation of pass $k$.
\end{proof}

Therefore, to every input permutation there must correspond
a unique combination of $N$ together with appropriate fixed
order (say in lexicographical order of subscripts) 
of elements of a partition as in (\ref{eq.M}).
How many such partitions are there?
Choosing $a$ elements out of an ordered list of $a+b$
elements divides the remainder into a sequence of $a+1$ 
possibly empty sublists. Hence
there are 
\begin{equation}\label{eq.DM}
 D(N) := 
 {{N+np-1} \choose {np-1}}
\end{equation}
distinct partitions of $N$
into $np$ ordered nonnegative integral summands $m_{i,k}$'s. 

Consider a subset $S$ of  $n!/2^n$  input permutations,
and let the maximum number
of inversions among them be $M$. Clearly, $M > 0$. Then,  overestimating the 
number of partitions involved,
$\sum_{N=0}^M D(N) \geq n!/2^n$ which implies $M D(M) \geq  n!/2^n$. 
Then,
\begin{equation}\label{eq.Mnp}
\log (M D(M)) \geq (\log n!) - n.
\end{equation}
We know that $M\leq pn^2$ since every $m_{i,k} \leq n$.
We have assumed
$p < n$. 
Hence, $\log M  < 3 \log n$. 
The standard estimate gives 
$\log n! = n \log n - O(n)$ for $n \rightarrow \infty$.
Estimate $\log D(M)$ by \footnote{
Use the following formula (\cite{LiVi93}, p. 10), 
\[ \log {a \choose b} = b \log \frac{a}{b} + (a-b) \log \frac{a}{a-b}
+ \frac{1}{2} \log \frac{a}{b(a-b)} + O(1) .\]
}
\[
 \log {{M+np-1} \choose {np-1}}  =   (np-1) \log \frac{M+np-1}{np-1}
 +M\log \frac{M+np-1}{M} 
 + \frac{1}{2} \log \frac{M+np-1}{(np-1)M} + O(1). 
\]
The second term in the right-hand side is bounded as\footnote{Use 
$e^a>(1+ \frac{a}{b})^b$ for all $a>0$ and positive integer $b$.}
\[ \log \left( 1+ \frac{np-1}{M} \right)^{M}
< \log e^{np-1} \]
for all positive $M$ and $np-1>0$. Since $0 < p < n$ and $1 \leq M \leq pn^2$,
\[
\frac{1}{2(np-1)} \log \frac{M+np-1}{(np-1)M} \rightarrow 0
\]
for $n \rightarrow \infty$. Therefore,
$\log D(M)$ is majorized asymptotically by
\[ A= (np-1) \left( \log \left(  \frac{M}{np-1} +1 \right)+ \log e \right) \]
for $n \rightarrow \infty$. Altogether, 
$A + \log M \geq n \log n - O(n)$. 
With $p \leq (\epsilon / \log e) \log n$
($0 < \epsilon < 1$), 
this can be rewritten as
\[ (np-1)  \log ( \frac{M}{np-1} +1) \geq (1 - \epsilon ) n \log n  - O(n), \]
and further as
\[ \log  ( \frac{M}{np-1} +1) \geq 
(\frac{1 - \epsilon}{p} )  \log n - O(\frac{1}{ p}) .\]
The righthand
side is positive and asymptotic to the 
first term for $n \rightarrow \infty$. Hence, 
\[ M = \Omega ( p n^{1+ (1 - \epsilon)/p} ) . \]
\end{proof}

\begin{theorem}\label{theo.main}
The average computation time (number of inversions, for $p= o(\log n)$, 
and comparisons,
for $n/2 \geq p= \Omega ( \log n)$) in 
$p$-pass Shellsort on lists of $n$ keys is at least
$
\Omega \left( pn^{1+1/p}
\right)$ 
for every increment sequence.
The average is taken with all
lists of $n$ items equally likely (uniform distribution).
\end{theorem}
\begin{proof}
Assume the terminology above. 
Take $S$ to be the special set of  $n!/2^n$  input permutations using
the {\em least} number of inversions. Then,
the number of inversions made by algorithm $A$ for every
permutation not in $S$ is at least $M$ in the previous
proof. The theorem follows,
since for $p=o(\log n)$ ($\epsilon (n) \rightarrow 0$ for $n \rightarrow \infty$
in Theorem~\ref{theo.shelllb}) the
lower bound on the expected number
of inversions of the sorting procedure is at least
\[ (1- \frac{1}{2^n}) \Omega ( p n^{1+ 1/p} ) + \frac{1}{2^n} \Omega(0) =
 \Omega ( p n^{1+ 1/p} ) ; \]
and for $p = \Omega ( \log n)$, the trivial lower bound on the number
of comparisons is vacuously $pn= \Omega (pn^{1+1/p})$.
\end{proof}

Our lower bound on the average-case  can be compared
with the Plaxton-Poonen-Suel 
$\Omega (n^{1 + \epsilon / \sqrt{p} } )$ worst case lower bound \cite{PPS92}.
Some special cases of the lower bound on the average-case complexity are:

\begin{enumerate} 
\item
For $p=1$ our lower bound is asymptotically tight: it is the
average number of inversions for Insertion Sort.

\item For $p=2$, Shellsort requires $\Omega (n^{3/2} )$ inversions
(the tight bound is known to be $\Theta (n^{5/3})$ \cite{Kn73});

\item For $p=3$, Shellsort requires $\Omega (n^{4/3} )$ inversions
(the best known upper bound is $O(n^{23/15})$ in \cite{JK96});

\item For $p= \log n / \log \log n$, Shellsort requires
$\Omega (n \log^2 n / \log \log n )$ inversions; 

\item For $p=\log n$, Shellsort requires $pn = \Omega (n \log n)$
comparisons. When we consider comparisons, 
this is of course the lower bound of average number of 
comparisons for every sorting algorithm.


\item In general, for $n/2 \geq p = p(n) \geq \log n$, 
Shellsort requires $\Omega (n \cdot p(n))$ comparisons
(it requires that many comparisons anyway
since every pass trivially makes about $n$ comparisons).

\end{enumerate} 
In \cite{Se97} it is mentioned that the existence of an increment
sequence yielding an average $O(n \log n)$ Shellsort has been
open for 30 years. The above lower bound on the average shows
that the number $p$ of passes of such an increment sequence (if it exists)
is precisely $p=\Theta (\log n)$; all
the other possibilities are ruled out.

\section{Conclusion} 

The average-case performance of Shellsort has been one of the most
fundamental and interesting open problems in the area of algorithm
analysis. 
The simple average-case analysis of
Insertion Sort (1-pass Shellsort), 
and similar analyses of Bubble sort, 
stack-sort and queue-sort are given
in the preliminary version of this paper \cite{JLV99} and serve
as further examples to demonstrate
the generality and simplicity of our 
technique in analyzing sorting algorithms in general.
Some open questions are:
\begin{enumerate}
\item
Tighten the average-case lower bound for Shellsort. Our bound is not
tight for $p = 2$ passes. 
\item
Is there an increment sequence for $\log n$-pass
Shellsort so that it runs in average-case $\Theta (n \log n)$?
\end{enumerate}

\section{Acknowledgements}
We thank Don Knuth, Ian Munro, Vaughan Pratt, and
Osamu Watanabe for discussions
and references on Shellsort.




\end{document}